\documentclass[twocolumn,showpacs,preprintnumbers,amsmath,amssymb,superscriptaddress,prb]{revtex4}

\usepackage{graphicx}
\usepackage{dcolumn}
\usepackage{bm}
\usepackage{multirow}%
\usepackage[latin1]{inputenc}
\usepackage[T1]{fontenc}

\usepackage{verbatim}
\usepackage{color}

\newcommand{\half}{\frac{1}{2}}

\newcommand{\bt}{BaTiO$_3$}
\newcommand{\pt}{PbTiO$_3$}

\begin{document}

\title{Computational study of (111) epitaxially strained
  ferroelectric perovskites \bt{} and \pt{}}
\author{Riku Oja}
\affiliation{Laboratory of Physics, Helsinki University of
  Technology, P.O. Box 1100, FI 02015 TKK, Finland}
\author{Karen Johnston}
\affiliation{Laboratory of Physics, Helsinki University of
  Technology, P.O. Box 1100, FI 02015 TKK, Finland}
\affiliation{Theory of Polymers, Max-Planck-Institute for
Polymer Research, P.O. Box 3148, D 55021 Mainz, Germany}
\author{Johannes Frantti}
\affiliation{Laboratory of Physics, Helsinki University of
  Technology, P.O. Box 1100, FI 02015 TKK, Finland}
\author{Risto M. Nieminen}
\affiliation{Laboratory of Physics, Helsinki University of
  Technology, P.O. Box 1100, FI 02015 TKK, Finland}

\begin{abstract}
The phase transition behaviour of {\pt} and {\bt} under (111) epitaxial strain is investigated using density-functional theory calculations.  From tensile strains of +0.015 to compressive strains of -0.015, {\pt} undergoes phase transitions from $C2$ through two $Cm$ phases and then to $R3m$.  The total polarisation is found to be almost independent of strain.  For the same range of strains {\bt} undergoes phase transitions from a single $Cm$ phase, through $R3m$ and then to $R\bar{3}m$.  In this case the application of compressive strain inhibits and then completely suppresses the polarisation on transition to the non-polar $R\bar{3}m$ phase.
\end{abstract}

\maketitle

\section{Introduction}

Perovskite materials have been extensively studied due to their
numerous structures and properties \cite{King-Smith1994a}.  At
room temperature bulk BaTiO$_3$ and
PbTiO$_3$ are ferroelectric with spontaneous polarizations
of around 25 $\mu {\rm C cm}^{-2}$  and 80 $\mu {\rm C cm}^{-2}$ ,
respectively.
Epitaxial strain can have a considerable effect on the structural and electrical properties of perovskite thin films \cite{choi2004,neaton2003a,grigoriev2008,Dieguez2004a}. The properties of thin films can, therefore, be fine-tuned by growing the films epitaxially and coherently on a substrate with a desired lattice constant.

The effect of strain has already been studied for perovskite thin
films oriented in the (001) direction on a cubic substrate
\cite{Pertsev1998a,Pertsev_2000a,bungaro,Dieguez2004a,Dieguez2005a}. (111) epitaxial films have different symmetries than (001) films and, therefore, the behaviour will also be different.
The temperature-strain phase diagram of \pt{} (111) films has been modelled using a thermodynamic mean-field approach \cite{Tagantsev2002a}.  For room temperature and below, it was predicted that the film should have $3m$
point symmetry for compressive strains and two different $m$ symmetry phases for tensile strains, with first-order transitions between the phases. In this paper we investigate the stability of various phases of (111) films of \pt{} and \bt{} under epitaxial strain at T=0~K using density-functional theory calculations. We show that the zero-temperature phases of \pt{} mostly agree with the higher-temperature results of Ref. \onlinecite{Tagantsev2002a}.  \bt{} exhibits a qualitatively different behaviour from \pt{} and, surprisingly, the application of compressive strain is found to suppress the polarization.

\section{Method}

The parent space group for a (111) epitaxially strained
perovskite film is $R\bar{3}m$. In this work we used 15-atom unit cells, but only considered symmetries compatible with a 5-atom primitive cell. Periodic boundary conditions were used, and the effect of the substrate is implicitly included by
constraining the $\hat{a}$ and $\hat{b}$ lattice parameters to be (1,0,0) and
($- \half$,$\frac{\sqrt{3}}{2}$,0) and scaling both lattice vectors with mismatch strain. We have not included a surface as a ferroelectric field in vacuum gives rise to a depolarizing field, which suppresses polarization. This approach allows us to separate the effect of epitaxial strain from surface effects.

\begin{figure}[ht!]
\includegraphics[width=0.3\textwidth]{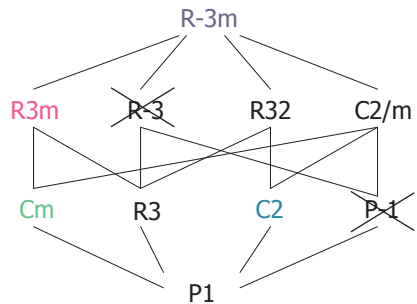}
\caption{\label{Rbar3m} (Color online) The subgroups of $R\bar{3}m$ with a possible 5-atom primitive cell. Crossed symmetries are not present in (111) epitaxial perovskite.}
\end{figure}

\begin{figure}[ht!]
\includegraphics[width=0.45\textwidth]{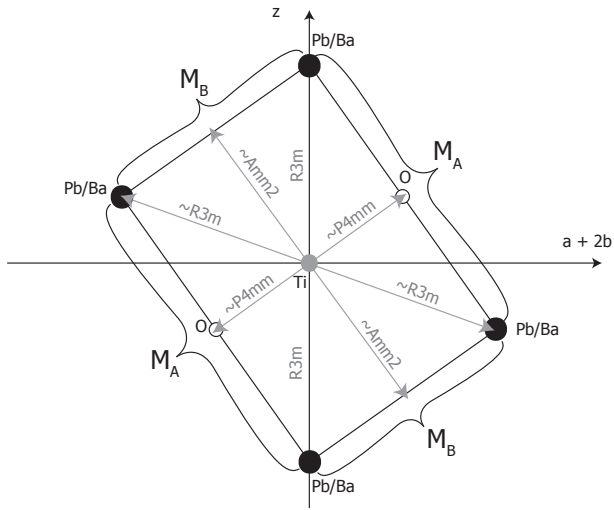}
\caption{\label{Cm} Possible polarization directions in the mirror plane of the $Cm$ symmetry. Tilde denotes the presence of symmetry at zero strain.}
\end{figure}

The ISOTROPY
code \cite{Isotropy} and the Bilbao Crystallographic Server \cite{bilbao} were used to find space groups compatible with
a perovskite (111) film. Fig.~\ref{Rbar3m} shows the
various subgroups of $R\bar{3}m$ with a possible 5-atom primitive cell and their symmetry relations. The monoclinic geometry of the substrate implies that $P\bar{1} $ will actually have a higher $C2/m$ symmetry, and $R\bar{3}$ actually has the higher $R\bar{3}m$ symmetry for perovskites as the mirror-plane breaking Wyckoff positions are not occupied. Therefore, eight possible space groups exist and are shown in Table~\ref{tab:sg}. The possible unstrained bulk symmetries that reduce to these space groups when (111) strain is applied are listed in Table \ref{tab:str}, with the $Cm$ phase divided in three types as shown in Fig. \ref{Cm}.

\begin{table}[ht!]
\begin{ruledtabular}\begin{tabular}{rllll}
\multicolumn{2}{c}{Space group} & $c$-axis         & Polarization & $N_P$  \\ \hline
166 &$R\bar{3}m$ & $c_3\hat{z}$                      & $0$ & 0 \\
155 &$R32$       & $c_3\hat{z}$                      & $0$ & 1  \\
 12 &$C2/m$      & $c_1\hat{a}+c_3\hat{z}$           & $0$ & 0* \\ \hline
160 &$R3m$       & $c_3\hat{z}$                      & $P_3\hat{z}$& 3  \\
146 &$R3$        & $c_3\hat{z}$                      & $P_3\hat{z}$ & 4  \\
  8 &$Cm$        & $c_2\frac{\hat{a}+2\hat{b}}{\sqrt{5}}+c_3\hat{z}$ & $P_2\frac{\hat{a}+2\hat{b}}{\sqrt{5}}+P_3\hat{z}$& 7* \\
  5 &$C2$        & $c_3\hat{z}$                      & $P_3\hat{z}$& 5* \\
  1 &$P1$        & $c_1\hat{a}+c_3\hat{z}$           & $P_1\hat{a}+P_3\hat{z}$& 12*\\
\end{tabular}
\end{ruledtabular}
\caption{\label{tab:sg} Description of the non-polar and polar (111) epitaxial perovskite symmetry groups
  with a possible 5-atom primitive cell. Vectors are given in the standard rectangular Cartesian basis ($\hat{a},\frac{\hat{a}+2\hat{b}}{\sqrt{5}},\hat{z}$). $N_P$ is the number of free parameters in atomic fractional coordinates when we restrict the primitive cell to 5 atoms, although the symmetries marked with * would allow a 15-atom primitive cell, with more free parameters in fractional coordinates.}
\end{table}

\begin{table}[ht!]
\begin{ruledtabular}\begin{tabular}{ll}
Strained & Unstrained \\ \hline
$R\bar{3}m$ & $Pm\bar{3}m$ \\
$R3m$       & $R3m$ \\
$R3$        & $R3$ \\
$Cm$ ($M_A$)& $Cm$ and $P4mm$ \\
$Cm$ ($M_B$)& $Cm$ and $Amm2$ \\
$Cm$ (between $M_A$ and $M_B$) & $R3m$ \\
$C2$        & $Amm2$ \\
$P1$        & $P1$ and $Pm$ \\
\end{tabular}
\end{ruledtabular}
\caption{\label{tab:str} Considered strained symmetries and bulk symmetries that reduce to them when (111) strain is applied.}
\end{table}

For the (111) perovskite, fewer polarization directions are allowed than in the (001) case. The non-polar $R32$ and $C2/m$ space groups are not expected to be seen in the (111) perovskite \cite{Tagantsev2002a}, and they have not been considered. For displacements strictly perpendicular to the plane, the polarization $P_3\hat{z}$ and the space group $R3m$ is obtained. $R3m$ is predicted to be stable for compressive strains in \pt{} \cite{Tagantsev2002a}. $R3$ is a subgroup of $R3m$ that has polarization in the same direction but additional displacements of atoms to break the mirror plane.

For displacements parallel to a lattice vector in the plane, the polarization $P_1\hat{a}$ and the space group $C2$ is obtained. The stability of such a \pt{} phase is considered uncertain in Ref. \onlinecite{Tagantsev2002a}, as it depends on the selection of model parameters.

For displacements parallel to $\hat{a}+2\hat{b}$ (or, equivalently, $\hat{a}-\hat{b}$ or $\hat{a}+2\hat{b}$) in the plane, the space group $Cm$ is obtained, in which the polarization is only constrained to lie in a mirror plane perpendicular to the epitaxial plane.  According to the Landau theory, $Cm$ is stable for tensile strains in \pt{} \cite{Tagantsev2002a} (when using the parameters from Ref. \onlinecite{haun1987}), and the transition $Cm \to R3m$ is first-order.  Due to symmetry considerations, transition $Cm \to C2$ also has to be first-order.

Within the monoclinic $Cm$ space group, there are three inequivalent possible energy minima, with isotropic phase transitions between them. They correspond to the bulk space groups $R3m$, $Amm2$ and $P4mm$ and pseudocubic bulk polarization directions (111), (110) and (100), respectively \cite{Tagantsev2002a}.  In Ref.~\onlinecite{vanderbilt2001a}, the stability of the monoclinic phase in bulk is studied with high-order Devonshire theory, and possible monoclinic $M_A$ and $M_B$ phases (along with $M_C$, which would have $Pm$ symmetry that is not possible in a (111) biaxially strained perovskite) originate in the eighth-order expansion of the thermodynamic potential.  Our $Amm2$ minimum corresponds to $M_B$ and $P4mm$ minimum to $M_A$, as displayed in Fig.~\ref{Cm}.

Finally, a triclinic phase with no symmetry constraints ($P1$) would require twelfth-order terms in Ref. \onlinecite{vanderbilt2001a} and is predicted to be unlikely in the bulk. No such phase is considered in Ref. \onlinecite{Tagantsev2002a} either, and it is left outside the present study.

In a strained film, there are nonzero stresses $\sigma_1$ and $\sigma_2$ (in the standard Cartesian axes) in the epitaxial direction. Therefore, the thermodynamic potential that is minimized is the elastic enthalpy

\begin{equation}
\label{enthalpy}
H = U + e_m(\sigma_1+\sigma_2),
\end{equation}

where $U$ is the internal energy and $e_m$ the in-plane strain relative to the relaxed lattice constant of the phase in question. The phase with lowest elastic enthalpy at given mismatch strain is the equilibrium state of a coherent epitaxial single-domain film. The internal energies and stresses for each phase as a function of strain are obtained from density-functional theory (DFT) calculations after relaxing atomic positions and unit cell parameters within the constraints of Table~\ref{tab:sg}.

The DFT calculations were performed using the
Vienna {\it ab-initio} simulation package (VASP)
\cite{Kresse1993a,Kresse1996b}.
The core states were represented using the PAW method
\cite{Blochl1994a,Kresse1999a} and the semicore Pb, Ba, and Ti
were treated as valence electrons. The planewave cutoff energy
was 700 eV. The exchange and correlation
energy was described using the local density approximation (LDA). The polarization of the phases was calculated using Berry phase formalism \cite{king-smith1993a,marsman}.

The hexagonal (monoclinic for $Cm$) conventional cell contained 15 atoms and consisted of three identical 5-atom primitive cells. A $\Gamma$-centred Brillouin zone mesh of
6$\times$6$\times$4 was used.

Ionic relaxations were performed for various $c/a$ ratios (for all space groups) and $c$ axis tilt angles (for $Cm$) to obtain the optimal cell shape and volume for each fixed in-plane strain. Relaxations were stopped when the Hellmann-Feynman forces on the atoms were below 0.01 $e$V/\AA{}.

\section{Results}

The relaxed DFT bulk cubic lattice values were 3.895 \AA{} for \pt{} and 3.955 \AA{} for \bt{}, and they were used as the unstrained lattice constants. The calculated $c/a$ ratios in the bulk tetragonal phase were 1.04 for \pt{} and 1.003 for \bt{}, and corresponding polarizations in the bulk tetragonal phase were 0.77 ${\rm C m}^{-2}$ for \pt{} and 0.21 ${\rm C m}^{-2}$ for \bt{}. As expected, these values are somewhat smaller than the experimental values, which is typical of the LDA approximation. The experimental cubic lattice constants are 3.97 \AA{} \cite{numericaldata} and 4.00 \AA{} \cite{ghosez1998a}, the tetragonal $c/a$ ratios are 1.07 and 1.01 and the polarizations are 0.80 ${\rm C m}^{-2}$ \cite{wu26} and 0.26 ${\rm C m}^{-2}$ \cite{wu28}, for \pt{} and \bt{}, respectively.

\begin{figure}[ht!]
\includegraphics[width=0.5\textwidth]{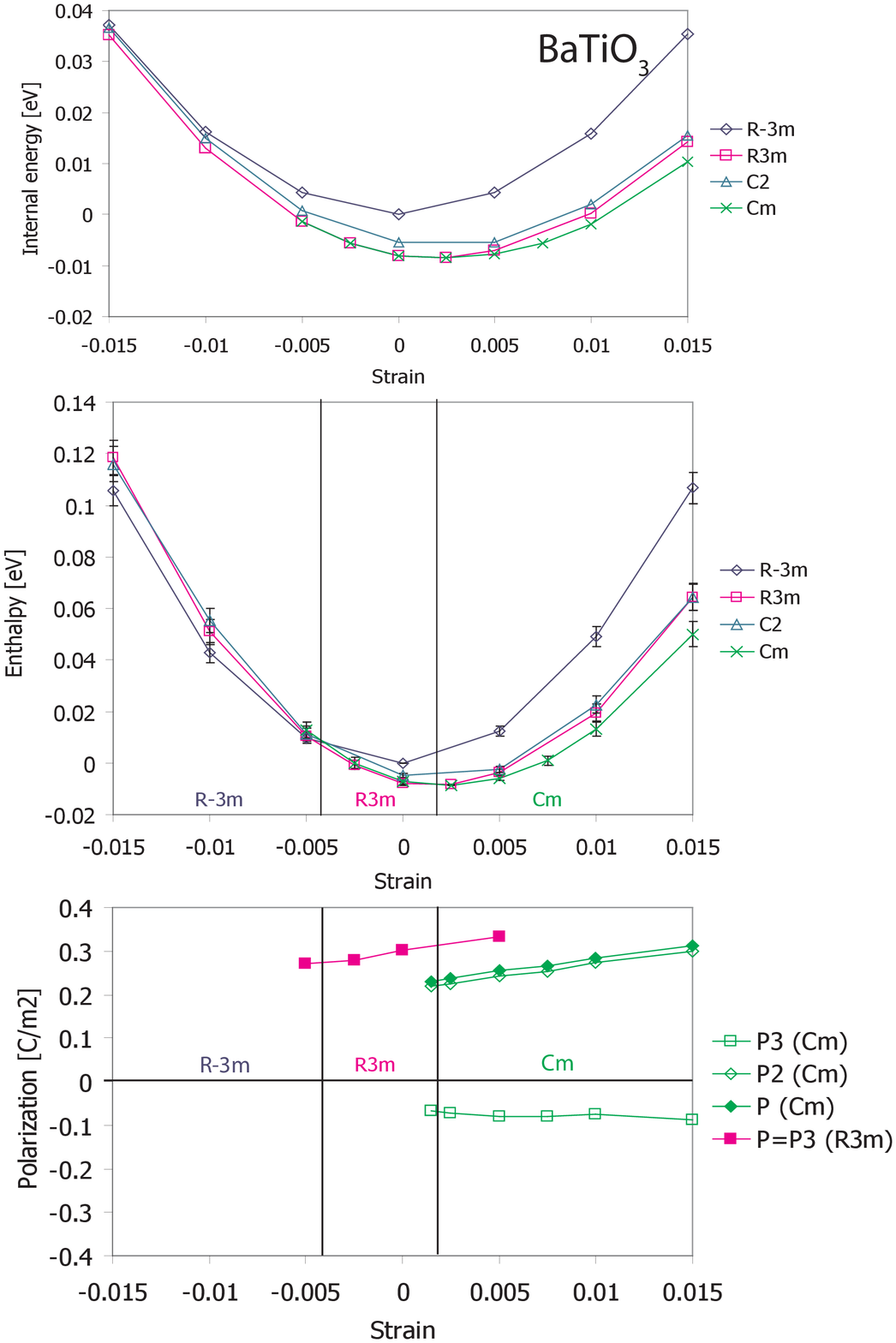}
\caption{\label{enenpol_bt} (Color online) (Top) The internal energies (per 5-atom primitive cell) given by VASP for considered phases of \bt{}. (Center) Elastic enthalpies (per 5-atom primitive cell) as given by Eq. \eqref{enthalpy} for considered phases of \bt{}. The error bars indicate approximate error in enthalpy due to stress inaccuracy in VASP and discretization in strain, $c/a$ ratio and $c$ axis angle. The vertical lines denote approximate points of phase transition. (Bottom) Polarization of the \bt{} phase with lowest elastic enthalpy at each strain. $P$ is the total polarization and $P_1$, $P_2$ and $P_3$ are the polarization components in the $\hat{a}$, $\frac{\hat{a}+2\hat{b}}{\sqrt{5}}$ and $\hat{z}$ directions, respectively. }
\end{figure}

\begin{figure}[ht!]
\includegraphics[width=0.5\textwidth]{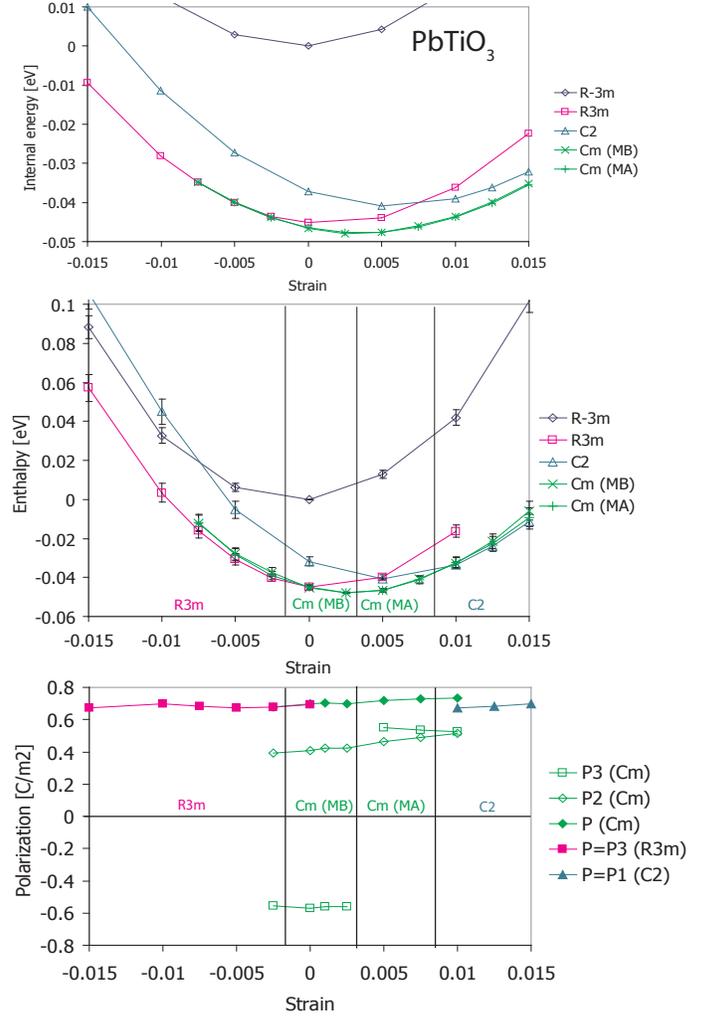}
\caption{\label{enenpol_pt} (Color online) (Top) The internal energies (per 5-atom primitive cell) given by VASP for considered phases of \pt{}. (Center) Elastic enthalpies (per 5-atom primitive cell) as given by Eq. \eqref{enthalpy} for considered phases of \pt{}. The error bars indicate approximate error in enthalpy due to stress inaccuracy in VASP and discretization in strain, $c/a$ ratio and $c$ axis angle. The vertical lines denote approximate points of phase transition. (Bottom)  Polarization of the \pt{} phase with lowest elastic enthalpy at each strain. $P$ is the total polarization and $P_1$, $P_2$ and $P_3$ are the polarization components in the $\hat{a}$, $\frac{\hat{a}+2\hat{b}}{\sqrt{5}}$ and $\hat{z}$ directions, respectively. }
\end{figure}

 The internal energies and corresponding elastic enthalpies for the considered epitaxial phases, with error estimates, are displayed as a function of strain in Fig.~\ref{enenpol_bt} for \bt{} and Fig.~\ref{enenpol_pt} for \pt{}. In the enthalpy curves, the energetically favored phases are denoted. The error in enthalpy is proportional to relative strain squared. For \pt{}, two separate but almost identical energy curves for the $Cm$ phases were obtained, corresponding to different energy minima within the same symmetry, by relaxing the unit cell starting from three different starting configurations.
For the strains considered, $R3$ was found to relax into the $R3m$ symmetry for both \pt{} and \bt{}, so no separate $R3$ curve is plotted.

In-plane ($P_1$ and $P_2$), perpendicular to plane ($P_3$), and total ($P$) polarization of the perovskites as a function of strain are displayed in Fig.~\ref{enenpol_bt} for \bt{} and Fig.~\ref{enenpol_pt} for \pt{}. The components displayed are strictly from the phase of lowest enthalpy for each strain as denoted, although the error due to discretization in strain, $c/a$ ratio and $c$ axis angle makes the preferred phase uncertain at large compressive or tensile strains and between the two \pt{} $Cm$ phases.

\subsection{BaTiO$_3$}

\begin{figure}[ht!]
\includegraphics[width=0.45\textwidth]{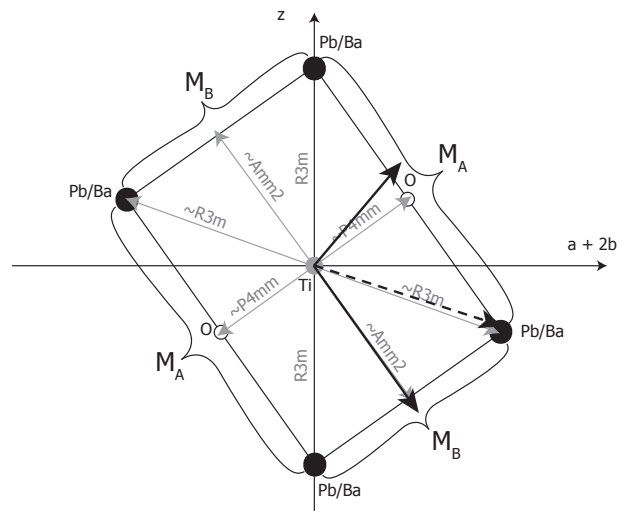}
\caption{\label{Cmptbt} Approximate polarization directions in the mirror plane of the $Cm$ symmetry. \bt{} adopts a single energy minimum: dashed arrow indicates polarization of \bt{} at tensile strain 0.0025. \pt{} adopts two energy minima: black arrows indicate polarization of \pt{} $M_B$ and $M_A$ phases at tensile strains 0 and 0.005, respectively. Tilde means symmetry direction at zero strain.}
\end{figure}

For \bt{}, only a single energy curve for the $Cm$ phase could be obtained, with relaxation to a single energy minimum instead of transitions between them. This energy minimum most closely corresponds to a nearly $R3m$ structure, with polarization towards a Ba atom as indicated in Fig.~\ref{Cmptbt}, and virtually no rotation of polarization with increasing strain. The result is in accordance with the known tendency of \bt{} to adopt the rhombohedral phase in the bulk. The transition $Cm \rightarrow R3m$ is seen to be first-order (with polarization vector as the order parameter), as predicted by an extension of Devonshire theory \cite{vanderbilt2001a}. The other \bt{} transition, ferroelectric-to-paraelectric ($R3m \rightarrow R\bar{3}m$) transition induced by compressive strain, is similarly seen to be first-order. Although the polarized $R3m$ phase has a lower internal energy than the $R\bar{3}m$ phase, the lower stresses observed in the paraelectric phase make it favored at compressive strains when elastic enthalpy is considered.

\subsection{PbTiO$_3$}

In \pt{}, the $Cm$ polarization direction at tensile strains is dominated by the tendency to adopt a nearly $P4mm$ ($M_A$) or $Amm2$ ($M_B$) structure. These polarization directions in the mirror plane of $Cm$ phase are sketched and labelled in Fig.~\ref{Cmptbt}. No relaxation to a nearly $R3m$ structure, as in \bt{}, is observed. Therefore, when the tensile strain is increased, a transition from $M_B$ to $M_A$ structure, with consequent flipping of one polarization component, is seen in Fig.~\ref{enenpol_pt}. Such a strain-induced transition is indeed predicted for (111) \pt{} in Ref. \onlinecite{Tagantsev2002a}, and all the \pt{} transitions are first-order (polarization vector changes discontinuously) as predicted.

This is in apparent contradiction with the fact that bulk \pt{} is known to adopt $P4mm$ symmetry when unstrained; however, the unstrained (111) epitaxial case is different from the bulk. Even at zero strain, the (111) substrate inhibits the tetragonal elongation of unit cell required for the $P4mm$ phase to be stable, and the $Amm2$ ($M_B$) geometry is adopted instead. When increasing tensile strain, an $M_A$ structure closer to $P4mm$ is adopted, as it allows the polarization to align more parallel to the strain.

\begin{table}
\begin{ruledtabular}
\begin{tabular}{lccc|c|c}
  Strain & -0.01 & 0 & 0.01 & \text{Cubic} & \text{Measured bulk (RT)}\\
\hline
 \pt{} & 1.84 & 1.77 & 1.71 & 1.48 & 3.4\cite{peng}\\
 \bt{} & 1.72 & 2.05 & 2.11 & 1.72 & 3.0\cite{prokopalo}\\
\end{tabular}
\end{ruledtabular}
\caption{Calculated LDA band gaps for lowest-enthalpy phases at different in-plane strains of (111) epitaxial perovskite. In the pseudocubic Brillouin zone, the band gaps are indirect $\Gamma$-X and $\Gamma$-R gaps for \pt{} and \bt{}, respectively. Energies are in eV.}
\label{tab:gaps}
\end{table}

The single enthalpy differences between the \pt{} $Cm$ structures are smaller than possible errors in enthalpy, but the enthalpy curves of Fig.~\ref{enenpol_pt} as a whole display slight differences between the $M_A$ and $M_B$ structures. Polarization in the two phases rotates very little in the narrow strain range where each phase is stable, though it would rotate significantly more if $Cm$ were stable at larger tensile strains. At large tensile strains, although the $Cm$ phase still has the lowest internal energy, the $C2$ phase is favored in terms of elastic enthalpy, as its relaxed lattice constant is higher than that of the $Cm$ phase. Therefore, large tensile strain (around 1 \%) causes \pt{} polarization to align along a lattice vector in the (111) plane; the structure is similar to the $Amm2$ phase in bulk, but with lower symmetry due to strain.

\subsection{Bandgaps}

Finally, the LDA bandgaps for the favored phases were calculated, to compare to LDA bandgaps of cubic BaTiO$_3$ and PbTiO$_3$. While LDA seriously underestimates bandgap, any strain-induced changes in the LDA bandgaps may indicate widening or closing of the band gap due to strain. The results for the two perovskites are shown in Table~\ref{tab:gaps}. The band structures and gaps are very similar to calculations on cubic perovskites \cite{King-Smith1994a}, and the $\Gamma$-X type of indirect gap in \pt{} agrees with optical measurements of the room temperature bulk \cite{peng}. It can be seen that all non-pseudocubic phases have the effect of widening the gap compared to the calculated bulk cubic values, but there are no clear trends or major changes in the size of the bandgap.

\section{Discussion and conclusions}

{\bt} has $Cm$ symmetry for large tensile strains and undergoes a phase transition to the $R3m$ phase at a strain of $\approx$0.002.
In both $R3m$ and $Cm$ phases, the total polarization decreases approximately linearly as compressive strain is increased.
The most surprising result is the adoption of the non-polar $R\bar{3}m$ phase below a strain of $-0.004$.
Therefore, compressive strain appears to inhibit and, finally, suppress, ferroelectricity in (111) epitaxial \bt. There are observations of such inhibition of remanent polarization in (111) epitaxial \bt{} on LaNiO$_3$-coated SrTiO$_3$ substrate \cite{zhu2006} which would inflict compressive strain due to lattice mismatch, and in (111) \bt{}/SrTiO$_3$ superlattices \cite{nakagawara2002}. However, it is also possible that a cell-doubling transition may occur, as only phases with a 5-atom primitive cell were considered in this study.

Both perovskites have a strain region in which the $R3m$ phase is stable and undergoes a first-order phase transition into the $Cm$ phase when tensile strain is increased. The geometry of the $Cm$ phase differs in the two perovskites, with a single minimum in \bt{} and first-order isomorphic transition between two $Cm$ phases in \pt{}, with switching of the vertical polarization component.  Such a transition was predicted in Ref.~\onlinecite{Tagantsev2002a} although their phase diagram has a lower limit of $-50^{\rm o}$~C and our first-principles calculations are valid at absolute zero.  Another indication of this isomorphic transition is the observation of  inversion of the vertical polarization in (111) PZT thin-film capacitors \cite{stolichnov2002}.  However, the strain in these thin-film capacitors is unknown.

In \pt{} it is observed that as tensile strain grows, the $C2$ phase (with in-plane polarization) is adopted.  Indeed, in Ref.~\onlinecite{Tagantsev2002a} it is noted that such a phase could be stable instead of $Cm$, depending on the parameters used in the thermodynamic expansion. The total polarization of the stable phases of \pt{} is observed to be almost independent of the applied strain.  The direction of polarization, but not its magnitude, changes with (111) strain.  Such a suppressed dependence of polarization on strain has been observed earlier in tetragonal (100) epitaxial PbZr$_{0.2}$Ti$_{0.8}$O$_3$ (PZT) \cite{lee}. This was proposed to be due to the fact that the polarisation of PZT is already large so the Pb ions are less sensitive to strain \cite{lee}.   Since {\pt} also has large atomic displacements it is likely to exhibit similar behaviour to PZT.

The magnitudes of the calculated polarization are similar to those observed in the (100) epitaxial case for both perovskites \cite{choi2004,lee}, apart from our predicted suppression of \bt{} polarization at compressive strain. Such suppression under compressive strain has also been observed in (111) epitaxial \bt{} \cite{zhu2006,nakagawara2002}, with remanent polarization of only 0.1 $\mu {\rm C cm}^{-2}$, opposed to 19.9 $\mu {\rm C cm}^{-2}$ in the (100) case \cite{zhu2006}. We are aware of no other reports on the magnitude of observed polarization in (111) epitaxial \pt{} or \bt{}.

The calculations presented here do have a number of limitations. First, the calculations give the internal energy at zero temperature.  A full strain-temperature phase diagram of a (111) epitaxial perovskite could be produced by parameterising an effective Hamiltonian model using first-principles results~\cite{Dieguez2004a,Dieguez2005a}. Second, the effect of the surface has not been considered. Third, no cell-doubling phases, such as octahedral rotations or other complex structures that require larger than 5-atom primitive cells, were considered, even though they may appear in perovskites in the stress regime considered here (e.g. Ref.~\onlinecite{frantti2007}).  The stability of the phases suggested here could be confirmed, for example, by calculating phonon modes. This was left beyond the scope of this study due to the large computational requirements of investigating each phase under various strains. Fourth, ionic zero-point motion has a significant effect on the adopted phases as a function of pressure at low temperatures, and it could significantly alter the phase diagram \cite{iniguez2002}.  For \bt{} in particular, the shallowness of the potential wells makes this effect considerable up to several hundred Kelvin in the GPa pressure range \cite{iniguez2002}, corresponding to strains considered here.  This effect could be taken into account by an effective Hamiltonian, if a path-integral quantum Monte Carlo technique, instead of classical Monte Carlo, were used in thermodynamic simulations. A final point to be noted is that while LDA is known to well reproduce many perovskite properties, the fact that it somewhat underestimates atomic displacements when compared to experiments could have an effect on the energetics of the various phases, as their elastic enthalpies are very close to each other.

\begin{acknowledgments}
This work was funded by the Academy of Finland FinNano program, project number 128229. CSC (the Finnish
IT Center for Science) provided the computing resources.
\end{acknowledgments}


\end{document}